\documentclass[preprint,showpacs,prb,superscriptaddress,aps,floatfix]{revtex4-1}
\usepackage{rotating}
\usepackage{amsmath}
\usepackage{color}
\usepackage{graphicx}
\usepackage{courier}
\usepackage{siunitx}
\usepackage[sort&compress]{natbib}
\usepackage{appendix}

\newcommand{\qq}{{\bf q}}

\newcommand{\kk}{{\bf k}}

\newcommand{\w}{\omega}

\newcommand{\be}{\begin{equation}}
\newcommand{\ee}{\end{equation}}
\newcommand{\bea}{\begin{eqnarray}}
\newcommand{\eea}{\end{eqnarray}}
\newcommand{\ben}{\begin{equation*}}
\newcommand{\een}{\end{equation*}}
\newcommand{\bean}{\begin{eqnarray*}}
\newcommand{\eean}{\end{eqnarray*}}
\renewcommand{\v}[1]{{\bf #1}}

\def\hbn{\textit{h}-BN}

\renewcommand{\thesection}{\Roman{section}}

\newcommand{\cinam}{CNRS/Aix-Marseille Universit\'e, Centre Interdisciplinaire de Nanoscience de Marseille UMR 7325 Campus de Luminy, 13288 Marseille cedex 9, France}
\newcommand{\torvergata}{University of Rome Tor Vergata, Rome, Italy}

\newcommand{\piim}{Universit\'e Aix-Marseille, Laboratoire de Physique des Interactions Ioniques et Moléculaires (PIIM), UMR CNRS 7345, F-13397 Marseille, France}
\newcommand{\etsf}{European Theoretical Spectroscopy Facilities (ETSF)}
\newcommand{\cambridge}{Cavendish Laboratory, University of Cambridge, J.\,J.\,Thomson Avenue, Cambridge CB3 0HE, United Kingdom}

\begin{document}
\title{Theory of phonon-assisted luminescence in solids: \\application to hexagonal boron nitride}
\author{E. Cannuccia} 
\affiliation{\torvergata}
\affiliation{\piim}
\author{B. Monserrat} 
\affiliation{\cambridge}
\author{C. Attaccalite}
\affiliation{\torvergata}
\affiliation{\cinam}
\affiliation{\etsf}

\date{\today}

\begin{abstract}
We study luminescence of hexagonal boron nitride (\hbn) by means of non-equilibrium Green's functions plus finite-difference electron-phonon coupling.
We derive a formula for light emission in solids in the limit of a weak excitation that includes perturbatively the contribution of electron-phonon coupling at the first order. This formula is applied to study luminescence in bulk \hbn. This material has attracted interest due to its strong luminescence in the ultraviolet region of the electromagnetic spectrum [Watanabe \textit{et al.}, Nature Mat. \textbf{3}, 404(2004)]. The origin of this intense luminescence signal has been widely discussed, but only recently a clear signature of phonon mediated light emission started emerging from the experiments [Cassabois \textit{et al.}, Nature Phot. \textbf{10}, 262(2016)]. By means of our new theoretical framework we provide a clear and full explanation of light emission in \hbn.

\end{abstract}           

\maketitle

\emph{Introduction.}
In standard solid state physics textbooks\,\cite{kittel1996introduction} direct band gap semiconductors are considered efficient light emitters while indirect ones are regarded as inefficient. 
Silicon is a typical example: its indirect nature prohibits applications in optoelectronics. This fact has motivated significant research activity to engineer silicon and transform it into a direct gap semiconductor by means of defects, nanostructuring, etc. A more recent and remarkable example of indirect to direct gap transition is represented by MoS$_2$ nanostructuring\,\cite{splendiani2010emerging}. The luminescence signal increases by four orders of magnitude passing from multi-layer to single layer MoS$_2$ with an associated indirect to direct band gap transition.  \\ 
Hexagonal boron nitride (\hbn) seems to defy this rule: it has a large indirect band gap of about 7 eV, but it has recently attracted much attention from the scientific community as a very efficient light emitter in the ultraviolet\,\cite{watanabe2004direct}. An internal quantum yield of $\simeq 45\%$ has been reported for \hbn, much closer to the $\simeq 50\%$ one of ZnO (direct band gap) than to the $0.1\%$ one of diamond (indirect band gap)\,\cite{schue2018direct}.
\begin{figure*}[ht]
\centering
\includegraphics[width=0.95\textwidth]{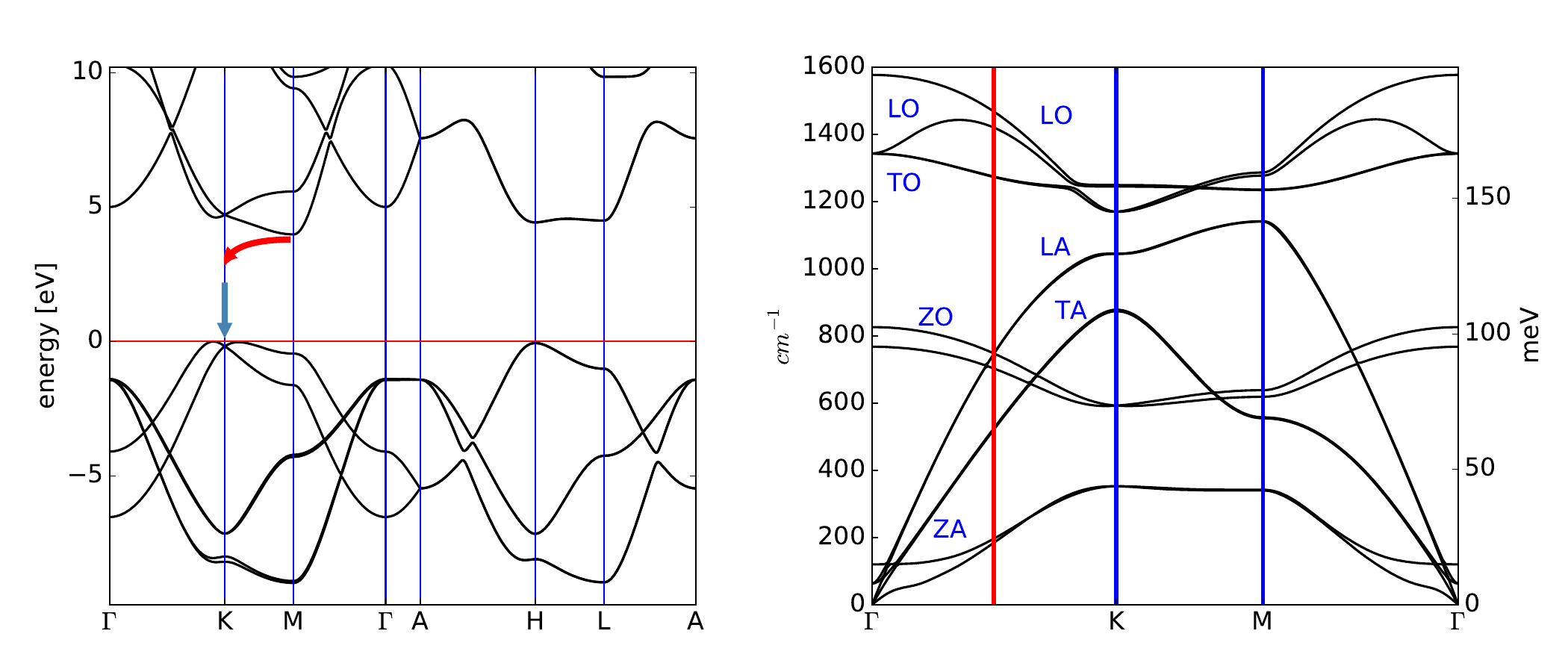}
     \caption{\footnotesize{[Color online] Panel $(a)$ the electronic band structure of \hbn. The red arrow represents the scattering with a phonon between $M$ and $K$ and the blue one the emission of a photon. Panel $(b)$ the phonon band structure of \hbn. The red vertical line marks the position of the phonon \textbf{q}-point involved in the luminescence process, see the red arrow in panel $(a)$. \label{phonons_and_bands}}}
\end{figure*}
This goes against the common wisdom that indirect band gap insulators are bad light emitters and in fact the strong luminescence signal was initially attributed to direct exciton recombination\,\cite{watanabe2004direct}. In order to shed some light on this phenomena different theories were proposed, including the presence of defects\,\cite{PhysRevB.83.144115} or  a dynamical Jahn-Teller effect\,\cite{PhysRevB.79.193104}. But only recently, thanks to more accurate and precise measurements, it has been shown that luminescence originates from phonon assisted recombination\,\cite{cassabois2016hexagonal}.
This interpretation has also been confirmed by studing  the isotopic effect\,\cite{vuong2018isotope} and polarization-resolved photoluminescence measurements\,\cite{vuong2016phonon}.
  In this manuscript we develop a new approach to study phonon-assisted luminescence that treats on the same footing electron-hole and electron-phonon interactions. We apply our new methodology to study \hbn~ luminescence, a subject that has remained puzzling until recent years.
  Since excitons play a crucial role in the optical response of \hbn, in order to describe luminescence we need a theory that includes both electron-hole interactions and electron-phonon scattering. \\
In the literature two important approaches have been proposed to study bulk luminescence, one based on non-equilibrium Green's functions\,\cite{hannewald2000theory} and another on density matrix theory\,\cite{kira2011semiconductor}. Both these approaches include the effect of the electron-hole interaction and the formation of excitons at different levels of approximation. 
In this work we follow Hannewald and co-workers\,\cite{hannewald2000theory} who formulate luminescence in terms of non-equilibrium Green's function theory. 
Inclusion of exciton-phonon coupling is more involved. Phonon-assisted luminescence has been studied by Kira and Koch using the polaron picture\,\cite{chernikov2012phonon,chernikov2012phonon}. Their approach has the advantage of being non-perturbative in the electron-phonon coupling while conserving the analytic structure of the zero-phonon response functions. Another possibility would be to include the electron-phonon diagrams directly in the Bethe-Salpeter equation\,\cite{antonius2017theory,ostreich1998coulomb}, but we discarded this option because of its complexity.\\
In recent years a new way has emerged to include the electron-phonon coupling by means of a finite differences approach\,\cite{monserrat2018electron}.
The approach has several advantages including the ability to combine it with any underlying electronic structure method and to include terms beyond the lowest order in the electron-phonon interaction\,\cite{monserrat2018electron}.
Finite difference methods have been used to study the effects of electron-phonon scattering in multiple contexts including optical absorption\,\cite{zacharias2015stochastic}, topological phase transitions\,\cite{monserrat2016topo}, and superconductivity\,\cite{kotliar2013superconduct}.
In all these studies electrons were always considered independent particles. Here, in order to include excitonic effects in the response functions we combine finite difference electron-phonon coupling calculations with Green's function theory and apply this novel method to the study of luminescence spectra.

\emph{Light emission from non-equilibrium Green's functions.}
Light absorption and emission can be described by means of Green's functions theory. Both these processes are related to the two-particle polarization functions. 
The different two-particle correlation functions can be obtained from the solution of the so-called Bethe-Salpeter equation (BSE) at equilibrium or out-of-equilibrium\,\cite{lucia}.
    In the BSE particle-hole pairs are coupled by correlation effects\,\cite{strinati}.
    In the literature different levels of approximation for this coupling have been presented, from the T-matrix\,\cite{kwong2009self} that can account for the effects of finite excitonic populations, to the simpler static ladder approximation\,\cite{lucia}. In this work we use the latter. The static ladder approximation has been shown to reproduce well the optical properties of \hbn~at equilibrium\,\cite{wirtz2006excitons}, and since we consider very few excited carriers we expect similar results to  hold also in our case.
    In this approximation the particle-hole coupling is composed of two terms, one deriving from the density fluctuations through the Hartree potential, called $v$, and the other one due to the fluctuations of the screened exchange potential, the so-called electron-hole interactions $W$. The last term is normally considered in a static approximation, and derived from the COHSEX self-energy\,\cite{lucia}.
\begin{figure*}[ht]
\centering
\includegraphics[width=0.9\textwidth]{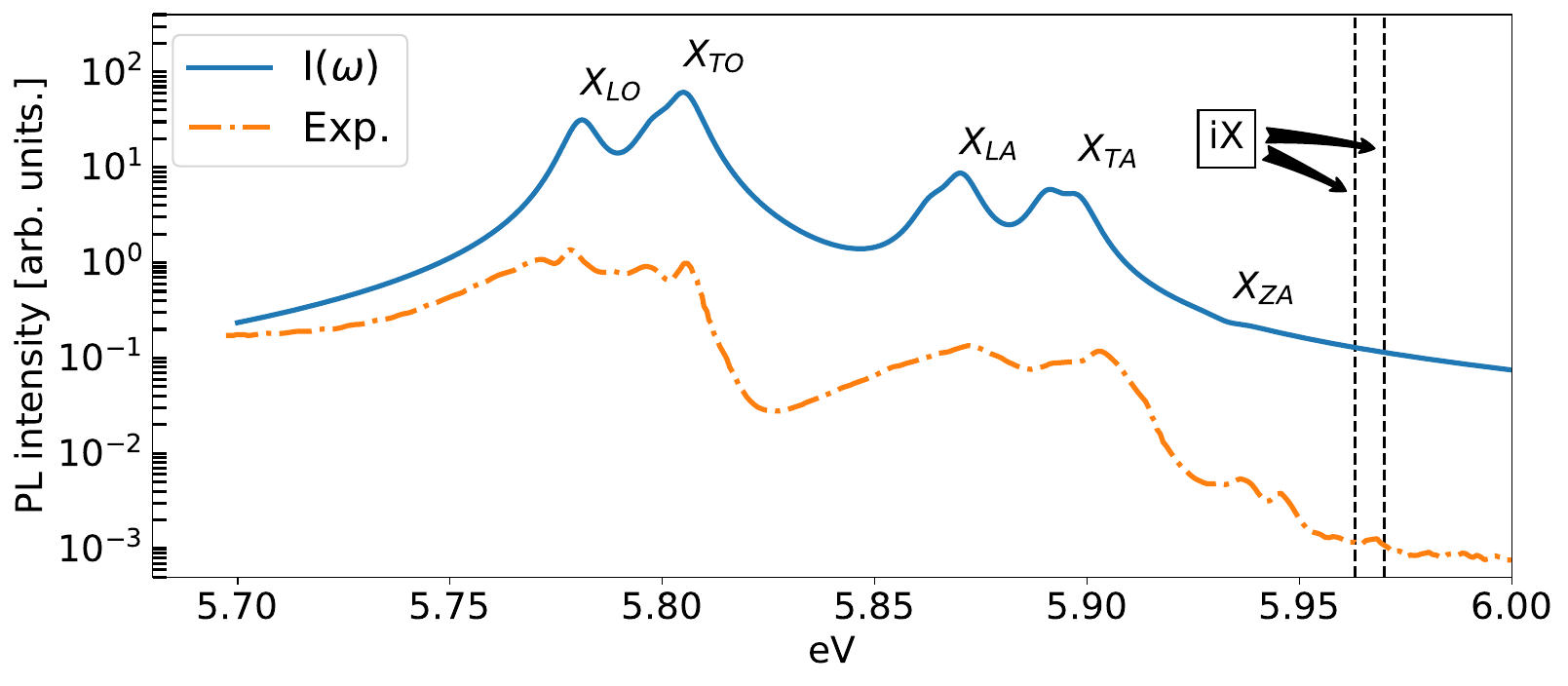}
    \caption{\footnotesize{[Color online] Blue continous line: luminescence spectra calculated using [Eq.~\ref{ihbb}] for bulk \hbn~. The vertical lines represent the position of the $iX$ doublet, that has zero dipole matrix elements. Orange dash-dotted line: experimental results taken Ref.\,\cite{cassabois2016hexagonal}. We choose the normalization of both spectra in such a way to have an optimal visualization in a single figure.} \label{plspectra}}
\end{figure*}
   In non-equilibrium conditions, when part of the carriers has been excited to the conduction bands, it is possible to write down an out-of-equilibrium BSE to describe optical absorption. This approach has been used to investigate excitonic Mott transitions\,\cite{Schleife2011} or transient absorption in pump-probe experiments\,\cite{hannewald2000theory}.
The out-of-equilibrium BSE then reads:
\be
 [\Delta \varepsilon_K \delta_{K,K'}+ \sqrt{f_{K}} \Xi_{K,K'}  \sqrt{f_{K'}} ] A^{K'}_\lambda= E_\lambda A^K_\lambda,
\label{eq:BSE_ne}
\ee
where $K$ is a general index for $\{vc\kk\}$, $f_K = f_{v\kk} -  f_{c\kk}$ are the occupation factors and
 $E_\lambda$ and  $A^K_\lambda$  are the eigenvalues and eigenvectors, namely the exciton energy and wave-functions. The $\Delta \varepsilon_K = \varepsilon^{QP}_{c\kk} -  \varepsilon^{QP}_{v\kk} $ are the quasi-particle energy differences. The kernel is written as $\Xi=i[W-v]$\,\cite{lucia}.
The screened Coulomb potential $W$ that appears in [Eq.~\ref{eq:BSE_ne}] is calculated taking into account the non-equilibrium electron and hole distributions.
In [Eq.~\ref{eq:BSE_ne}] we use the symmetrization via $\sqrt{f_K}$, introduced by Schleife and co-workers\,\cite{Schleife2011}, to ensure that the Bethe-Salpeter Hamiltonian remains pseudo--Hermitian even in the presence of fractional occupations, or Hermitian in the case of the Tamm-Dancoff approximation.\\
Notice that the reality of the solution of [Eq.~\ref{eq:BSE_ne}]
is no longer guaranteed in stationary excited states. However, if the excitation is weak, namely the quasi-particle occupations only slightly differ from their equilibrium values, as in the case studied here, then the solutions continue to be real\,\cite{perfetto2016first}.
  The solution of [Eq.~\ref{eq:BSE_ne}] leads to the different response functions that are related to the absorption or emission spectra\,\cite{hannewald2000theory}. \\
Following the derivation of Refs.\,\cite{hannewald2000theory,melothesis}, in the limit of low excitations, within the Tamm-Dancoff approximation, the luminescence power spectra can be expressed as:
\be
I(\w)  \propto  \sum_{\lambda} | \Pi_{\lambda} |^2 f^{<}_{\lambda} \delta{(\w-E_{\lambda})},
\label{eq:BSE_I}
\ee
where  $E_\lambda$ are the eigenvalues of the out-of-equilibrium BSE [Eq.~\ref{eq:BSE_ne}], $\Pi_\lambda$ are the exciton dipole matrix elements, and $f^<_\lambda$ their occupations. For a definition of $\Pi_\lambda$, $f^<_\lambda$ and the derivation of [Eq.~\ref{eq:BSE_I}] see Sec.~2 of the Supplementary Information (SI).\\
The excitonic occupation $f^<_\lambda$ is often approximated with a Bose-Einstein distribution $f^<_\lambda \simeq n_B(E_\lambda,T)$ in light emission\,\cite{torun2018inter,schue2018direct} while it is equal to one in the absorption process at equilibrum.
As one can see from [Eq.~\ref{eq:BSE_I}], the luminescence spectra resonates at the same frequencies as the absorption does but excitons are weighted in a different way. \\ 
Notice that in the independent particle approximation, light emission is described by an equation equivalent to [Eq.~\ref{eq:BSE_I}] where the sum over excitons is replaced by one over independent electron-hole pairs.\\
\emph{Scattering with phonons.}
In order to include the coupling between electrons and phonons, we consider the electron-phonon interaction at first order in the atomic displacements:
\be
\hat H_{el-ph} = \frac{1}{\sqrt{N_p}}\!\! \sum_{\substack{\kk,\qq \\ m n \nu }} \!g_{mn\nu}(\kk,\qq) \,
      \hat c_{m\kk+\qq}^\dagger \hat c_{n\kk}(\hat a_{\qq\nu}+\hat a_{-\qq\nu}^\dagger) 
      \label{eq:epc}
\ee
where $\hat c^\dagger_{m\kk}/\hat c_{n\kk}$ ($\hat a^\dagger_{-\qq\nu}/\hat a_{\qq\nu}$) are the fermionic (bosonic) creation/annihilation operators with crystal momentum $\kk$ ($\qq$) and $N_p$~is the number of unit cells in the Born-von K\'arm\'an supercell. \\
The electron-phonon Hamiltonian [Eq.~\ref{eq:epc}] can be treated in time dependent perturbation theory in the independent particle basis\,\cite{bassani1975electronic} or the excitonic one\,\cite{perebeinos}.
In the case of indirect semiconductors the main effect of [Eq.~\ref{eq:epc}] is to make active the transitions to finite-$\qq$ excitons, which are otherwise inactive in the optical absorption/emission processes. The theory of phonon-assisted indirect optical transitions was developed by Hall, Bardeen, and Blatt\,\cite{hall1954infrared}, and employed in first-principles calculations in Ref.\,\cite{noffsinger2012phonon}.
In our work we extend their approach to light emission by means of time-dependent perturbation theory in the excitonic basis\,\cite{perebeinos}.
We consider the adiabatic limit for the dipole matrix elements, while we retain dynamical effects only in transition energies. Then we express the first order correction to the dipole matrix elements as a derivative with respect to a phonon mode, as shown in Ref.\,\cite{zacharias2015stochastic}, and we get the final formula for phonon-assisted luminescence as:
\bea
I^{\mathrm{BSE}}(\omega; T)&\propto&  \sum_{\mu \lambda} \frac{\partial^2| \Pi_{\lambda} |^2}{\partial x^2_{\nu}} f^{<}_{\lambda}  \left [  \delta{(\w-E_{\lambda} - \omega_{\nu})}  \frac{n_B(\omega_{\nu}; T)}{2 \omega_{\nu}}  + \right . \nonumber \\
&+& \left . \delta{(\w-E_{\lambda} + \omega_{\nu})} \frac{1 + n_B(\omega_{\nu}; T) }{2 \omega_{\nu}} \right ], 
\label{ihbb}
\eea
where $n_B$ is the Bose distribution function and $x_\nu$ is the atomic displacement associated to the phonon mode with frequency $\omega_\nu$.
Notice that in [Eq.~\ref{ihbb}] there should also be a term describing light emission without phonons, but in the case of indirect gap insulators this term vanishes. For more details of the derivation of Eq.~\eqref{ihbb} see Secs.~II and III in the SI. The two terms in [Eq.~\ref{ihbb}] describe the absorption and emission of a phonon respectively. The first process is less important at low temperatures due to the low number of available phonons.\\ 
In the adiabatic approximation ($\omega_\nu=0$ in the delta functions) this formula recovers the standard static approximation, similar to the one derived in Ref.\,\cite{zacharias2016one} for the dielectric constant.
Notice that [Eq.~\ref{ihbb}], and in general the HBB theory, do not capture the smooth increase of the absorption coefficient with temperature, and the concurrent redshift of the absorption onset, but they include dynamical effects that are crucial to reproduce luminescence spectra.\\ 
Eq.~\ref{ihbb} is the central formula of this work, and it will be used in the following to evaluate phonon-assisted luminescence from first principles.\\
\emph{Computational details.}
We studied \hbn\, by means of Density Functional Theory with the {\tt Quantum Espresso} code\,\cite{pwscf}, and Many-Body Perturbation Theory with the {\tt Yambo} code\,\cite{yambo}. The lattice parameters in our simulations are $a= 2.5 \buildrel _\circ \over {\mathrm{A}}$ and $c/a=2.6$\,\cite{solozhenko_ssc1995}. For the calculation of the density we used a $12 \times 12 \times 4 $ k-points grid, a cutoff of $80$~Ry for the wave-functions and the LDA approximation to the exchange correlation functional. The same parameters have been used for the phonon dispersion calculation.\\
In bulk \hbn\, the optical response originates from the two $\pi$ and two $\pi^*$ bands and the gap is indirect between a point very close to $K$ (valence bands) and $M$ (conduction bands)\,\cite{sponza2018exciton}. Therefore in the Bethe-Salpeter equation [Eq.~\ref{eq:BSE_ne}] we include only 2 valence and 2 conduction bands and the corresponding number of bands in the supercells. A scissor operator of 2.328~eV has been applied to the Kohn-Sham band structure to reproduce the correct position of the lowest exciton with momentum $\qq = M -K$\,\cite{gwcorrection}, also called $iX$ exciton in the literature. The dielectric constant that enters the electron-hole interaction of [Eq.~\ref{eq:BSE_ne}] has been calculated using $40$ bands and a $4$~Ha cutoff.\\ 
In order to study phonon-assisted emission we build a nondiagonal supercell\,\cite{lloyd2015lattice} 
such that the $K$ and $M$ points are mapped to $\Gamma$, for more details see Sec.~I in the SI. In this way the phonon modes with momentum $ \qq = K - M$ are folded to the $\Gamma$-point in the supercell. 
Then the derivatives with respect to the phonon modes that appear in [Eq.~\ref{ihbb}] can be calculated by finite differences\,\cite{monserrat2018electron}. A $3$-point formula is used to evaluate the second derivative, and only the phonon modes compatible with the $\qq = K-M $ vector have been included in the calculations. We introduce a broadening of $0.0045$~eV in the luminescence spectra in order to simulate the experimental one.
Finally we take a density of excited carriers of $n~= 10^{15}$~cm$^{-3}$ between $K$ and $M$\,\cite{schue2016dimensionality}  to evaluate the luminescence spectra and consider the emission wave vector parallel to the \emph{c} axis. \\
\emph{Results.}
In Fig.~\ref{phonons_and_bands}, panel $(a)$, we report the electronic band structure of \hbn\, calculated at the DFT level, and in panel $(b)$ the phonon band structure. The maximum of the valence band is located close to the $K$ point while the minimum of the conduction bands is at $M$. This situation is typical of other layered materials and it is due to the interlayer interaction that induces a transition from direct to indirect band gap going from monolayer to bulk. 
The inclusion of correlation effects does not change this picture, as verified by numerical simulations\,\cite{sponza2018exciton} and experimental measurements\,\cite{schuster2018direct}.
Direct light emission is forbidden in \hbn\, due to the indirect band gap, and only phonon-assisted luminescence is allowed.
In the full excitonic dispersion there are two minima (at two $\qq$-points which fall very close to the $\qq=M-K$ point) and both can contribute to the luminescence\,\cite{sponza2018exciton}. We here consider the $iX$ exciton for the reasons highlighted below.  
Since the top of the valence band falls very close to the $K$ point of the Brillouin zone, we approximate this point as the $K$ one. In this way the supercell containing both $K$ and $M$ is large but still computationally treatable, and the evaluation of electron-phonon coupling requires a feasible computational cost. 
The phonons involved in the luminescence process are those with a momentum compatible with the $\qq = K-M $ vector, being the ones to contribute to the momentum balance between the bottom of the conduction band and the top of the valence band.
Therefore they fall in the middle of the Brillouin zone between $\Gamma$ and $K$, and are reported in panel $(b)$ of Fig.~\ref{phonons_and_bands} with a red line.\\ 
In order to study luminescence we first diagonalize the Bether-Salpeter Hamiltonian in the supercell without including electron-phonon coupling, so as to find the position of the $iX$ exciton. We found that the $iX$ is formed by two dark excitons, separated of about 0.01~eV (dashed lines in Fig.~\ref{plspectra}). Then we include electron-phonon coupling by means of [Eq.~\ref{ihbb}]. The two $iX$ excitons are replicated by the different phonon modes and aquire a finite optical weight as shown in the final spectra in Fig.~\ref{plspectra}.\\
\emph{Comparison with experiments.}
In Fig.~\ref{plspectra} we report both the luminescence spectra calculated with our method and the experimental results from Cassabois and co-workers\,\cite{cassabois2016hexagonal}. The two spectra compare very well.  We correctly reproduce the position and intensity of the main peaks. The doublets measured in the luminescence spectra are generated by the LO-TO (LA-TA) splitting along the $\Gamma - K $ line.
We also report the position of the $iX$ exciton that is not visible in luminescence, even if in the experiments there is a small peak probably due to the presence of impurities.
The overtones, visible as lower intensity redshifted sidebands in experimental luminescence spectra\,\cite{vuong2017overtones} are not reproducible in our calculations\,\footnote{Notice that in our computational approach the electron-phonon scattering is limited to one phonon at a time and so we cannot reproduce the overtones involving interlayer shear modes and the asymmetric tails of the emission peaks\,\cite{vuong2017overtones}.}. 
The splitting of the $X_{TO}$, $X_{TA}$ and $X_{LA}$ peaks originates from the choice of supercell and it is related to the intrinsic exciton structure in our calculations.
 

In order to estimate the relative intensity of phonon-assisted luminescence in~\hbn, we calculate the ratio between a hypothetical direct emission in \hbn~ and the phonon-mediated one (more details are given in Sec.~IV of the SI). 
We find the ratio between direct and indirect emission to be $I^{\mathrm{DIR}}/I^{\mathrm{IND}} \simeq 10^2$, implying that in going from bulk to monolayer \hbn~ the luminescence will increase, but not as much as in the MoS$_2$ case\,\cite{splendiani2010emerging}. Our result is also in agreement with the finding of Schu\'e and co-workers\,\cite{schue2016dimensionality} that measured an increase of the ratio between direct and indirect peaks as the number of \hbn~layers decreases.\\ 
\emph{Summary and conclusions.}
    In this work we study phonon-assisted luminescence by means of a new non-equilibrium Green's functions based formula plus time-dependent parturbation theory in the exciton space and then the electron-phonon coupling matrix elements are evaluated by a finite differences method. 
We find that luminescence in \hbn\, is dominated by phonon-assisted transitions and that its intensity is unexpectedly large when compared with direct transitions.\\
\emph{Acknowledgments.} 
The research leading to these results has  received funding from the European Union Seventh Framework Program under grant agreement no. 785219 Graphene Core2. E.C.\ acknowledges support from the Programma per Giovani Ricercatori - 2014 ``Rita Levi Montalcini''. B.M. acknowledges support from the Winton Programme for the Physics of Sustainability, and from Robinson College, Cambridge, and the Cambridge Philosophical Society for a Henslow Research Fellowship.  C.A. acknowledges PRACE for computational resources on Marconi at CINECA (Grant No. Pra16\_4181L), A. Zappelli for the management of the computer cluster \emph{Rosa}, and D. Sangalli, F. Kusmartsev, E. Perfetto, G. Stefanucci, F. Ducastelle, H. Amara, A. Plaud and L. Sponza for useful discussions.  

\newpage

\appendixpage
\begin{appendices}
\renewcommand{\thesection}{\Roman{section}}

\section{Non-diagonal supercell generation}

A procedure for the construction of the smallest nondiagonal supercell commensurate with a given $\mathbf{q}$-point in the vibrational Brillouin zone (BZ) was recently described in Ref.\,\cite{lloyd2015lattice}. In the present paper we extend that procedure to construct the smallest nondiagonal supercell commensurate with a pair of $\mathbf{q}$-points, $(\mathbf{q}_1,\mathbf{q}_2)$. 

For a single $\mathbf{q}$-point of fractional reduced coordinates $(m_1/n_1,m_2/n_2,m_3/n_3)$, the smallest nondiagonal supercell has a size equal to the lowest common multiple (LCM) of $(n_1,n_2,n_3)$\,\cite{lloyd2015lattice}. We make the ansatz that for a pair of $\mathbf{q}$-points of fractional reduced coordinates $(m_1/n_1,m_2/n_2,m_3/n_3)$ and $(p_1/r_1,p_2/r_2,p_3/r_3)$ with associated LCM $l_1$ for $(n_1,n_2,n_3)$ and $l_2$ for $(r_1,r_2,r_3)$, then the smallest nondiagonal supercell has a size of at most $l_1\times l_2$. In practice, we iterate over all supercells of sizes smaller or equal to $l_1\times l_2$ until we find a supercell commensurate with both $\mathbf{q}$-points. 

\begin{figure}[ht]
\centering
\includegraphics[width=0.25\textwidth]{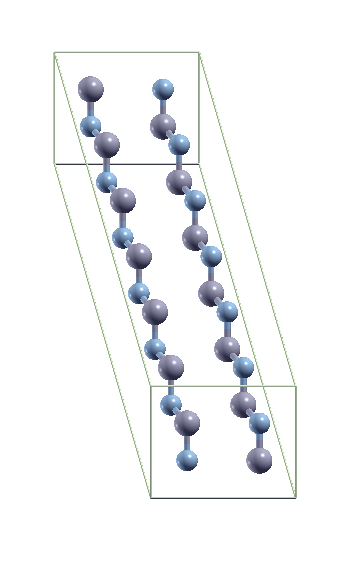}
     \caption{\footnotesize{[Color online] Smallest supercell where the phonon modes with momentum  $\mathbf{q}=K-M$ are mapped at the $\Gamma$ point, boron atoms in azure and nitrogen ones in blue-violet. \label{supercell}}}
\end{figure}

In the present paper we are interested in a supercell that contains both $K=(1/3,1/3,0)$ and $M=(0,1/2,0)$ $\mathbf{q}$-points. The smallest such supercell has $6$ primitive cells, and the associated supercell matrix is given by:
\begin{equation}
\begin{pmatrix}
1 & 2 & 0 \\
0 & 6 & 0 \\
0 & 0 & 1
\end{pmatrix}.
\end{equation}
This is the supercell used in the main text to capture the effects of the phonon modes with momentum $\mathbf{q}=K-M$.
The supercell contains 24 atoms and it is reported in Fig.~\ref{supercell}. 
 \\
Note that in the full BZ there are 6 different $\qq$ vectors that connect each point $K$ with the different $M$ points, and not all of them are included in our supercell. However these q-points are equivalent by symmetry and therefore their contribution to the final spectra only acts as a multiplicative factor. 

\section{Luminescence from non-equilibrium Green's functions}

In non-equilibrium conditions, when part of the carriers have been excited to the conduction bands, it is possible to write down an out-of-equilibrium Bethe-Salpeter Equation (BSE) to describe optical properties\,\cite{Schleife2011,hannewald2000theory,PhysRevB.63.075202,de2016unified}. The validity of this approach is discussed in detail in Ref.\,\cite{perfetto2015nonequilibrium}.
The out-of-equilibrium BSE reads:
\be
 [\Delta \varepsilon_K \delta_{K,K'}+ \sqrt{f_{K}} \Xi_{K,K'}  \sqrt{f_{K'}} ] A^{K'}_\lambda= E_\lambda A^K_\lambda.
\label{eq:BSE_ne2}
\ee
where $K$ is a general index for $\{vc\kk\}$, $f_K = f_{v\kk} -  f_{c\kk}$ are the occupation factors and
 $E_\lambda$ and  $A^K_\lambda$  are the eigenvalues and eigenvectors, namely the exciton energies and wave-functions. $\Delta \varepsilon_K = \varepsilon^{QP}_{c\kk} -  \varepsilon^{QP}_{v\kk} $ are the quasi-particle energy differences. The kernel is written as $\Xi=i[W-v]$\,\cite{lucia}.
The screened Coulomb potential $W$ that appears in [Eq.~\eqref{eq:BSE_ne2}] is calculated taking into account the non-equilibrium electron and hole distributions.
In [Eq.~\eqref{eq:BSE_ne2}] we use the symmetrization via $\sqrt{f_K}$, introduced by Schleife and co-workers\,\cite{Schleife2011}, to ensure that the Bethe-Salpeter Hamiltonian remains pseudo--Hermitian even in the presence of fractional occupations, or Hermitian in the case of the Tamm-Dancoff approximation. The same symmetrization is used for the Green's functions:
\be
L_{K,K'} = \sqrt{f_K} \tilde L_{K,K'} \sqrt{f_{K'}}.
\label{symL}
\ee
  From the solution of [Eq.~\eqref{eq:BSE_ne}] it is possible to construct the different response functions that are related to the absorption or emission spectra\,\cite{hannewald2000theory}. In particular, the power spectrum, that is related to the luminescence, is written in terms of the lesser two-particle Green's function as\,\cite{hannewald2000theory,melothesis,enderlein1979general,henneberger1996quantum}:
\be
I(\w)  \propto  \sum_{K,K'} \Im \left[  \Pi^*_{K} L^{<}_{K,K'} (\w)  \Pi_{K'} \right ]
\label{pspectrum}
\ee
where $\Pi_K = \langle \psi_{v\kk} | \hat j | \psi_{c\kk}\rangle$ are the current operator matrix elements in the single particle orbitals basis, and $L^{<}_{K,K'}(\w)$ is the lesser Green's function defined as:
\be
L^{<}_{K,K'}(\w) = \left [ 1 - L^{0 r}(\w) \Xi \right ]^{-1}_{K,K''} L^{0<}_{K''}(\w) \left [ 1 - \Xi  L^{0 a}(\w) \right ]^{-1}_{K'',K'}. 
\ee
$L^{0 r/a}$ are the non-interacting two-particle advanced/retarded Green's functions and  $L^{0 <}$ is the non-interacting lesser one:
\bea
L^{0 <}_K &=& 2 i \pi f^{<}_K \delta (\w -\Delta \varepsilon_K ), \\
L^{0 r/a}_K &=& i \frac{f_K}{\w - \Delta \varepsilon_K \pm i \eta},
\eea
where  $f^{<}_K = f_{c \kk} (1 - f_{v \kk} )$ and $\eta$ is a positive infinitesimal. The advanced and retarded Green's functions obey the Dyson equations:
\bea
\left [ 1 - \tilde L^{0r}  \Xi \right ]_{K,K'} &=& \tilde L_K^{0r} \tilde M^r_{K,K'} \label{dysonlar1}, \\
\left [ 1 - \Xi \tilde L^{0a}  \right ]_{K,K'} &=& \tilde M^a_{K,K'} \tilde L_{K'}^{0a}.  \label{dysonlar2}
\eea
Using the solutions of [Eqs.~\eqref{dysonlar1},\eqref{dysonlar2}] and the symmetrized version of the Green's functions [Eq.~\eqref{symL}] we can write $L^{<}_{K,K_2}$ as:
\be
L^{<}_{K,K_2}(\w) = \sqrt{f_K} \left (\tilde M^{r} \right )^{-1}_{K,K_1} \left( \tilde L^{0r}\right)^{-1}_{K_1} \frac{ L^{0<}_{K_1}}{f_{K_1}} \left( \tilde L^{0a}\right)^{-1}_{K_1} \left (\tilde M^{a} \right )^{-1}_{K_1,K_2} \sqrt{f_{K_2}}. 
\ee
The matrices  $\tilde M^{a/r}_{K,K_1}(\w)$ can be written in the excitonic basis, obtained from the solution of [Eq.~\eqref{eq:BSE_ne}].  Within the Tamm-Dancoff approximation, these matrices are Hermitian and they read:
\be
\left [  \tilde M^{r/a}(\w) \right]^{-1}_{K,K'} = i\sum_\lambda \frac{\langle K | \lambda \rangle \langle \lambda | K' \rangle}{\w - E_\lambda \pm i\eta }.
\ee
With these definitions we can now write $L^{<}_{K,K_2}$ in the excitonic basis too:
\be
L^{<}_{K,K_2}(\w) =2i\frac{\eta}{\pi} \sum_{\lambda_1, \lambda_2, K_1} \sqrt{f_K} \frac{\langle K | \lambda_1 \rangle \langle \lambda_1 | K_1  \rangle}{\w - E_{\lambda_1} + i\eta } \frac{f^<_{K_1}}{ f_{K_1}}\frac{\langle K_1 | \lambda_2 \rangle \langle \lambda_2 | K_2 \rangle}{\w - E_{\lambda_2} - i\eta }   \sqrt{f_{K_2}}.
\ee
In the limit of small $\eta \rightarrow 0$ we assume that the dominant contributions are those for $\lambda_1 = \lambda_2$, thus neglecting the remaining part $\lambda_1 \neq \lambda_2$. Then we get:
\be
L^{<}_{K,K_2}(\w) =2i\sum_{\lambda,K_1} \sqrt{f_K}  \langle K | \lambda \rangle \langle \lambda | K_1 \rangle \frac{f^<_{K_1}}{ f_{K_1}} \langle K_1 | \lambda \rangle \langle \lambda | K_2 \rangle  \sqrt{f_{K_2}} \delta (\w - E_{\lambda}).
\ee
In the limit of a small number of excited carriers, the occupations $f_K$ are different from zero only for valence-conduction transitions and can be approximated by the equilibrium ones that are constant with respect to the $K$ index. Then the formula for $L^{<}_{K,K_2}$ simplifies to:
\be
L^{<}_{K,K_2}(\w) =2i \sum_{\lambda,K_1} \langle K | \lambda \rangle \langle \lambda | K_1 \rangle f^<_{K_1} \langle K_1 | \lambda \rangle \langle \lambda | K_2 \rangle \delta (\w - E_{\lambda}).
\label{lsimple}
\ee
Now we define the lesser occupation and the dipole matrix elements in the excitonic basis set as:
\bea
f_\lambda^{<} &=& \sum_{K_1}  \langle \lambda | K_1  \rangle f^<_{K_1} \langle K_1 | \lambda \rangle \label{fexc}, \\
\Pi_\lambda &=& \sum_{K_1} \Pi_{K_1} \langle K_1 | \lambda \rangle.
\label{piexc}
\eea
Using [Eq.~\eqref{fexc}] and [Eq.~\eqref{piexc}] in [Eq.~\eqref{lsimple}], and also the definition of the power spectrum [Eq.~\eqref{pspectrum}] we finally get:
\be
I(\w) \propto \sum_{\lambda} | \Pi_{\lambda} |^2 f^{<}_{\lambda} \delta{(\w-E_{\lambda})}=\sum_{\lambda} | 
      \langle GS | \hat \Pi | \lambda \rangle |^2 f^{<}_{\lambda} \delta{(\w-E_{\lambda})}.
\label{simpleps}
\ee
where $|GS\rangle$ is the ground state without excitons.
This is the formula used in our manuscript to calculate luminescence. Notice that in the limit of a small number of excited carriers the out-equilibrium BSE [Eq.~\eqref{eq:BSE_ne}] reduces to the standard BSE\,\cite{strinati,lucia}, and in particular for the $h$-BN case we find no differences between the solution of the two equations due to the low density of excited carriers\,\cite{schue2018direct}. In the next section we describe how 
this formula is corrected perturbatively in the presence of phonons.

\section{Electron-phonon coupling and luminescence}
A well-established perturbative treatment of the phonon-sideband problem
was already developed in the 1970s\,\cite{rashba1982excitons} and it has been widely used in the literature until the present day\,\cite{perebeinos,perebeinos2008phonon}.
We start from the the electron-phonon Hamiltonian at first order in the atomic displacements:
\be
\hat H_{el-ph} = \frac{1}{\sqrt{N_p}}\sum_{\substack{\kk, \qq \\ m n \nu }} \!g_{mn\nu}(\kk,\qq) \,
      \hat c_{m\kk+\qq}^\dagger \hat c_{n\kk}(\hat a_{\qq,\nu}+\hat a_{-\qq,\nu}^\dagger)
      \label{eq:epc2}
\ee
where $\hat c^\dagger_{m\kk}/\hat c_{n\kk}$ ($\hat a^\dagger_{-\qq \nu}/\hat a_{\qq \nu}$) are the fermionic (bosonic) creation/annihilation operators with crystal momentum $\kk$ ($\qq$), $g_{mn\nu}(\kk, \qq)$ are the electron-phonon coupling matrix elements and $N_p$ is the number of unit cells in the Born-von K\'arm\'an supercell. Later we will consider only phonons at zero momentum because the relevant phonon modes at finite $\qq$ will be mapped to the $\Gamma$ point by means of a supercell approach (see Sec.~I above). \\
Our goal is to find a perturbative correction to the luminescence formula [Eq.~\eqref{simpleps}] induced by the interaction with phonons [Eq.~\eqref{eq:epc2}]. We consider the exciton wave-function, the solution of [Eq.~\eqref{eq:BSE_ne}]:
\be
| \Psi^\lambda_\qq \rangle =  \sum_{n,m,\kk} A_\lambda^{n,m,\kk }c^\dagger_{n,\kk+\qq} c_{m,\kk} | GS \rangle  
\ee
where $n$, $m$ are all possible electron-hole pairs and $A_\lambda^{n,m,\kk }$ are the eigenvectors of the BSE Hamiltonian. The $A_\lambda^{n,m,\kk }$ satisfy the orthogonality relation $\sum_{\lambda} A_\lambda^{n,m,\kk } A_\lambda^{n',m',\kk' } = \delta_{\kk,\kk'} \delta_{n,n'} \delta_{m,m'}$. 
The electron-phonon coupling matrix elements in the two-particle basis are
\be
g^{\nu}_{vc,\kk+\qq ; v'c',\kk} = g_{cc'\nu}(\kk,\qq)\delta_{vv'} - g_{v'v\nu}(\kk,\qq)\delta_{cc'}.
\ee
Note that when we project the electron-phonon interaction in the particle-hole basis we neglect non-excitonic terms such as the scattering by phonons of an electron from the conduction to the valence, and viceversa\,\cite{antonius2017theory}. In this case these terms are negligible because of the large gap of~\hbn. Rotating this matrix into the excitonic basis we can define the exciton-phonon matrix elements as
\be
B^{\lambda, \lambda'}_{\qq,\nu} = \sum_{\substack{v,c,\kk+\qq\\v',c',\kk}} A_{\lambda}^{*~v,c,\kk+\qq} g^{\nu}_{vc, \kk+\qq ; v'c', \kk} A_{\lambda'}^{v',c',\kk } 
\ee 

In the excitonic basis set the Hamiltonian [Eq.~\eqref{eq:epc}] reads: 
\be
\hat H_{ex-ph} = \frac{1}{\sqrt{N_p}} \sum_{\substack{\qq,\lambda, \lambda', \nu }} B^{\lambda, \lambda'}_{\qq,\nu} c^{\dagger}_{\lambda,\qq} c_{\lambda} (\hat a_{\qq \nu}+\hat a_{-\qq \nu}^\dagger),
\ee
where $c^{\dagger}_{\lambda}$ and $c_{\lambda}$ are exciton creation and annihilation operators.
In the presence of exciton-phonon coupling, the new excitonic wave-functions can be evaluated using time-dependent perturbation theory:
\bea
| \tilde \Psi^\lambda_0 \rangle = | \Psi^\lambda_0 \rangle+ \frac{1}{\sqrt{N_p}}  \sum_{\substack{\lambda' \neq \lambda \\ \qq, \mu}}\left \{ \frac{B^{\lambda, \lambda'}_{\qq,\nu} a_{\qq,\nu}}{E^{\lambda'}_\qq +  \w_{\qq,\nu} - E_0^\lambda}| \Psi^{\lambda'}_\qq \rangle  e^{-i(E^{\lambda'}_\qq +  \w_{\qq,\nu} - E_0^\lambda)t}+  \frac{B^{\lambda, \lambda'}_{-\qq,\nu}a^{\dagger}_{-\qq,\nu} }{E^{\lambda'}_\qq -  \w_{-\qq,\nu} - E_0^\lambda} | \Psi^{\lambda'}_{-\qq} \rangle e^{-i(E^{\lambda'}_\qq - \w_{\qq,\nu} - E_0^\lambda)t}\right   \}.
\nonumber
\eea
Then, in the spectral range of indirect emission (absorption) processes, we consider $\Pi_\lambda =0$ and we get for the emission spectra:
\bea
I(\omega; T)&\propto& \frac{1}{\sqrt{N_p}} \sum_{\substack{\qq, \lambda, \nu}} \left | \sum_{\lambda'\neq\lambda} \frac{B^{\lambda, \lambda'}_{\qq,\nu} \Pi_{\lambda'}}{(E^{\lambda'}_\qq +  \w_{\qq,\nu} - E_0^\lambda)} \right | ^2\left( 1 + n_B[\omega_{\qq \nu}; T] \right)f^{<}_\lambda \delta{(\w-E_{\qq}^{\lambda} + \omega_{\qq \nu})}  \nonumber \\
 &+&
\frac{1}{\sqrt{N_p}} \sum_{\substack{\qq, \lambda, \nu}} \left | \sum_{\lambda'\neq\lambda} \frac{B^{\lambda, \lambda'}_{-\qq,\nu} \Pi_{\lambda'}}{(E^{\lambda'}_\qq -  \w_{-\qq,\mu} - E_0^\lambda)}  \right | ^2
n_B[\omega_{\qq \nu}; T] f^{<}_\lambda \delta{(\w-E_\qq^{\lambda} - \omega_{-\qq \nu})}, 
\label{fullI}
\eea
where $n_B(\w)$ is the Bose function. The two terms in [Eq.~\eqref{fullI}] correspond to the processes of absorption and emission of a phonon, respectively. Notice that this result is equivalent to the Hall, Bardeen and Blatt theory\,\cite{hall1954infrared,bassani1975electronic}, with the difference that working in the excitonic basis allows us to include the electron-hole interaction. In principle we should also consider the effect of the electron-phonon coupling on the exciton distribution function $f^<_\lambda$, but this correction is of higher order and can be neglected. \\
Now we make a static approximation in the denominator or [Eq.~\eqref{fullI}] by setting $\w_{\qq,\mu} = 0$, while we retain $\w_{\qq,\mu}$ in the delta functions in order to maintain the correct peak energy positions. Then we consider [Eq.~\eqref{fullI}] in a supercell where only $\qq=0$ phonons are present and we get:
\bea
I(\omega; T)&\propto&  \sum_{\lambda \nu}  \left |  \sum_{\lambda' \neq \lambda} \frac{B^{\lambda, \lambda'}_{\nu} \Pi_{\lambda'}}{E^{\lambda'} - E^\lambda}  \right  |^2 \left (1 + n_B[\omega_{\nu}; T]\right) f^{<}_\lambda \delta{(\w-E_{\lambda} + \omega_{\nu})}  \nonumber \\
 &+& \sum_{ \lambda \nu}  \left | \sum_{\lambda'\neq\lambda}   \frac{B^{\lambda, \lambda'}_{\nu} \Pi_{\lambda'}}{E^{\lambda'} - E^\lambda}  \right |^2 n_B[\omega_{\nu}; T]f^{<}_\lambda \delta{(\w-E_{\lambda} - \omega_{\nu})}. 
\label{fullI_gamma}
\eea

Following Refs.\,\cite{zacharias2015stochastic,zacharias2016one}, we can map the perturbative expansion in term of derivatives with respect to the phonon modes:
\be
\frac{1}{2 \omega_{\nu}}\frac{\partial^2| \Pi_{\lambda} |^2}{\partial x^2_{\nu}}  \simeq  \left | \sum_{\lambda'\neq\lambda} \frac{B^{\lambda, \lambda'}_{\nu} \Pi_{\lambda'}}{E^{\lambda'} - E^\lambda}  \right |^2.
\ee
The two terms above are not exactly equal because the derivative of the  dipole matrix elements contains non-excitonic terms, but these are negligible in large gap insulators.\\ 
Substituting the above equation in [Eq.~\eqref{fullI_gamma}] we get the central formula used in the main text:
\bea
I^{\mathrm{BSE}}(\omega; T)&\propto&  \sum_{\lambda, \nu} \frac{\partial^2| \Pi_{\lambda} |^2}{\partial x^2_{\nu}} f^{<}_{\lambda}  \left [  \delta{(\w-E_{\lambda} - \omega_{\nu})}  \frac{n_B(\omega_{\nu}; T)}{2 \omega_{\nu}}  +  \delta{(\w-E_{\lambda} + \omega_{\nu})} \frac{1 + n_B(\omega_{\nu}; T) }{2 \omega_{\nu}} \right ].
\label{ibse}
\eea
The phonon frequencies $\w_\nu$ are usually large on the scale of thermal energies therefore only few optical phonons will be
present even at higher temperatures. Hence, pronounced phonon sidebands can only be found on the low-frequency side of the zero-phonon line in most luminescence experiments. For this reason, many studies omit the part of [Eq.~\eqref{ihbb}] related to the absorption of a phonon.\\

    In principle [Eq.~\eqref{ibse}] and [Eq.~\eqref{fullI}] require a fine sampling of the exciton and phonon dispersions around the $\qq = K-M$, where there are phonons and excitons compatible with momentum conservation. This would require huge supercells, as for example simply doubling or tripling the supercell size will not generate $\qq$ points close enough to K-M where exciton and phonon dispersions are important. As shown in the experiments of Cassabois and co-workers\,\cite{cassabois2016hexagonal}, the finite exciton and phonon dispersions are responsible for the different broadening of the peaks. We compensate this effect by introducing a finite broadening in the spectra, as discussed in the main text. In order to take into account these dispersions, one could interpolate the electron-phonon coupling\,\cite{PhysRevB.97.235146} or include it in the Elliott theory, as recently done for Germanium\,\cite{menendez2018phonon}. At present we are working to move our methodology in these directions.\\

Finally, in order to compare our results with previous calculations, we consider the limit of zero electron-hole interactions in [Eq.~\eqref{ibse}]. In this case, the Bethe-Salpeter equation becomes diagonal and the sum over exciton states can be replaced by electron-hole excitations. [Eq.~\eqref{ibse}] then reads:
\be
I^{\mathrm{IPA}}(\omega; T)\propto  \sum_{\substack{v,c,\kk \\ \nu}} \frac{\partial^2| p_{vc\kk} |^2}{\partial x^2_{\nu}} f_{c \kk} (1 - f_{v \kk} )   \left [  \delta{(\w-\Delta\varepsilon_{vc\kk}- \omega_{\nu})}  \frac{n_B(\omega_{\nu}; T)}{2 \omega_{\nu}}  +  \delta{(\w-\Delta\varepsilon_{vc\kk} + \omega_{\nu})} \frac{1 + n_B(\omega_{\nu}; T) }{2 \omega_{\nu}} \right ].
\ee
where $\Delta \varepsilon_{vc\kk} = \epsilon_{c\kk} - \epsilon_{v\kk}$. Now we can express the derivative of the momentum operator in terms of electron-phonon matrix elements using Eq.~15 of Ref.\,\cite{zacharias2015stochastic}. Except for the electronic occupations that are typical of light emission, we recover the formula used by Noffsinger and co-workers\,\cite{noffsinger2012phonon} for the phonon-assisted absorption when the adiabatic limit in the dipole matrix elements is taken.

\section{Intensity of the direct/indirect emission}
Moving from bulk \hbn~to a single layer, the system undergoes a transition from an indirect to a direct band structure\,\cite{wickramaratne2018monolayer}. This behaviour is typical of two-dimensional semiconductors\,\cite{Kang2016}. In this transition, luminescence becomes more intense due to the direct nature of the single-layer. This is in agreement with the measurement of Schue and co-workers\,\cite{schue2016dimensionality}, that report an increase of the intensity of the direct emission peak with respect to the indirect one when going from the bulk to a few layers\,\cite{paleari2018excitons}. In order to give a rough estimate of the intensity change in this transition, we evaluate the ratio between the strongest phonon-assisted peak obtained by [Eq.~\eqref{ibse}] and the first optical active excitons $\lambda'$ in bulk \hbn~:
\be
I^{\mathrm{DIR}}/I^{\mathrm{IND}} \simeq \frac{|\Pi_{\lambda'}|^2}{ \frac{1}{2\omega_\nu}\frac{\partial^2| \Pi_{\lambda} |^2}{\partial x^2_{\nu}} }.
\ee
This ratio yields a value of the order of $10^2$, meaning that luminescence from a single layer, where direct excitons are dominant, could be very strong compared to the bulk one, but in general we expect the effect to be smaller than in the MoS$_2$ case. 

\end{appendices}

\addcontentsline{toc}{chapter}{Bibliography}
\bibliographystyle{apsrev4-1}
\bibliography{phassisted}

\end{document}